\documentclass[preprintnumbers,amsmath,amssymb,twocolumn,showpacs]{revtex4}
\usepackage{graphicx}% Include figure files
\usepackage{dcolumn}% Align table columns on decimal point
\usepackage{bm}% bold math

\begin{document}

\title{Electromagnetically Induced Transparency from a Single Atom in Free Space}

\author{L. Slodi\v{c}ka$^1$, G. H\'etet$^{1,2}$, S. Gerber$^1$, M. Hennrich$^1$, and R. Blatt$^{1,2}$}

\affiliation{
$^1$ Institute for Experimental Physics, University of Innsbruck, A-6020 Innsbruck, Austria \\
$^2$ Institute for Quantum Optics and Quantum Information of the
Austrian Academy of Sciences, A-6020 Innsbruck, Austria}

\begin{abstract}
In this letter, we report an absorption spectroscopy experiment
and the observation of electromagnetically induced transparency
from a single trapped atom. We focus a weak and narrowband
Gaussian light beam onto an optically cooled $^{138}$Ba$^{+}$ ion
using a high numerical aperture lens. Extinction of this beam is
observed with measured values of up to 1.3\%. We demonstrate
electromagnetically induced transparency of the ion by tuning a
strong control beam over a two-photon resonance in a three-level
$\Lambda$-type system. The probe beam extinction is inhibited by
more than 75\% due to population trapping.
\end{abstract}

\maketitle

Atom-photon interfaces will be essential building blocks in future
quantum networks \cite{Cir97,DLCZ}. Here, photons are usually
adopted as the messengers due to their robustness in preserving
quantum information during propagation, while atoms are used to
store the information in stationary nodes. The efficient transfer
of quantum information between atoms and photons is then essential
and requires controlled photon absorption with a very high
probability. The requisite strong coupling can be achieved, for
example, using high finesse cavities \cite{Bru94,Pin00,Hoo00} or
large atomic ensembles \cite{Pol04,Phi01}, which are the most
studied routes towards such goals.

Coupling of radiation to a single atom in free space is generally
considered to be weak, however, technological advances, as
nowadays available with large aperture lenses \cite{Sor07} and
mirrors \cite{Son07}, recently led to reconsider this point of
view. Novel experiments demonstrated extinctions of about 10\%
from single Rubidium atoms \cite{Tey08}, single molecules
\cite{Zum08,Wri08} and quantum dots \cite{Vam07}. More
recently, a light phase shift of one degree was observed by tuning
an off-resonant laser to a single Rubidium atom \cite{Alj09}, and
non-linear switching was demonstrated with a single molecule
\cite{Hwa09}. These experiments demonstrate first steps towards
quantum optical logic gates and quantum memories with single atoms
in free space.

Long term and controlled storage of quantum information will
likely require electromagnetically induced transparency (EIT).
This technique is widely used to control the absorption of weak
light pulses or single photons in atomic ensembles
\cite{Phi01,Eis05} and in high-finesse cavities
\cite{Boo07}. Here, a two-photon Raman transition in lambda-type
three-level atoms is driven by the weak probe light together with
a strong control laser. The control laser leads to splitting of
the excited state by the AC Stark effect, which suppresses the
absorption of the resonant probe light. Consequently, the change
of the control laser intensity can gate the propagating probe
field between absorption and transmission. Furthermore, adiabatic
switching of the control light can trigger the storage and
retrieval of a probe photon onto and from the long-lived atomic ground states
\cite{Cir97,Fle05}.

So far, EIT has been a phenomenon specific to optically thick
media consisting of ensembles of many atoms \cite{Fle05}, where
both the optical fields and the atomic states are modified.
However, quantum information processing requires single
well-defined qubits, e.g. single atoms, that can be individually
manipulated to perform deterministic quantum gates. It thus
appears necessary to use single atom-single photons interactions
to distribute information over the nodes of a quantum network. In
single atom experiments, the related effect of coherent population
trapping has been observed on the fluorescence field, which
reveals modifications of the atomic population, but leaves the
transmitted optical fields without significant change. While EIT with a single atom in a cavity has been demonstrated just recently \cite{Mue10,Kam10,Abd10},  its free-space counterpart still remains to be proven. A system based on this technique could
easily be used as an efficient single atom switch or as a
programmable phase shifter of a weak coherent beam and/or a single
photon field in a quantum network.

Currently, trapped ions are widely investigated as one of the most
promising techniques for quantum information processing
\cite{Hae08}. Furthermore, the good control over the electronic
and motional states of ions in Paul traps makes them ideal systems
to investigate the coupling of radiation to single absorbers. So
far, with single ions spatially localized to a few nanometers in
free space, only absorption efficiencies of about 10$^{-5}$ to
10$^{-6}$ have been observed \cite{Win87,Sch10}.
In this work we investigate a first step towards a free-space
quantum interface by demonstrating an extinction of 1.3\% and
electromagnetically induced transparency from a single trapped
ion.
\begin{figure}[t]
\hspace*{0mm}\centerline{\scalebox{0.125}{\includegraphics{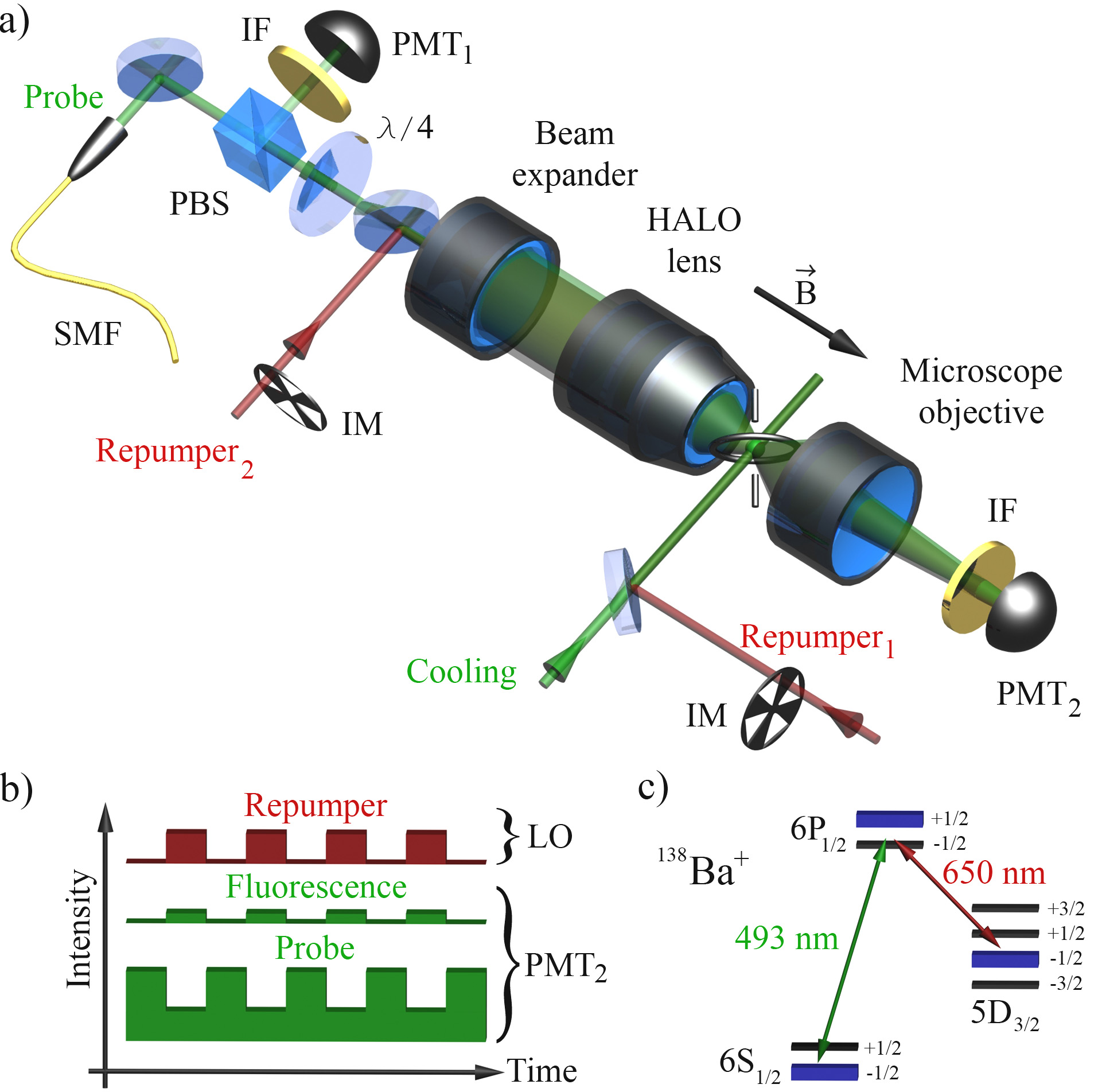}}}
\caption{a) Schematic of the experimental setup. PBS:~polarizing
beam splitter, $\lambda/4$:~quarter-wave plate, IF:~interference
filter to select the green fluorescence, IM:~intensity modulation,
PMT:~photomultiplier and SMF:~single mode fiber. The ion is cooled
by the {\it cooling} beam, while the {\it repumper no.\,1}
recycles population from the D state manifold to the S$_{1/2}$ to
P$_{1/2}$ transition. The {\it probe} beam is spatially adjusted
to match part of the atomic dipole emission profile and is
detected on the photomultiplier PMT$_2$. In the EIT experiment,
the probe field itself serves as a cooling beam and a
co-propagating {\it repumper no.\,2} is used instead. b) Detection
scheme. The probe beam and fluorescence modulation signals from
PMT$_2$ can be unambiguously distinguished due to their mutual
$\pi$ phase shift. The modulated signal from PMT$_2$ is mixed down
to DC and further analyzed, see text for details. c) A weak
magnetic field $\vec{\rm B}$ lifts the energetic degeneracy of the
Zeeman substates and creates an eight-level system. The levels
marked as bold lines were employed in the EIT experiment.}
\label{setup}
\end{figure}
First, we show a simple theoretical description of
extinction/reflection of a weak probe from a single atom. It uses
a perturbative input-output formalism to relate the incoming
field, $\hat{a}_{\rm in}$, and the outgoing field, $\hat{a}_{\rm
out}$, through their interaction with the atom \cite{Koc94}. In
the Markov limit this gives the relation $\hat{a}_{\rm
out}(t)=\hat{a}_{\rm in}(t) + i\sqrt{2\gamma_{\rm
in}}\hat{\sigma}(t)$, where $\hat{\sigma}(t)$ is the atomic
coherence and $\gamma_{\rm in}$ is the effective coupling of the
input to the atom. $\gamma_{\rm in}$ can also be expressed by the
total decay rate of the excited state $\gamma$ and the fraction
$\epsilon$ of the full solid angle covered by the incoming field
as $\gamma_{\rm in}=\epsilon\gamma$. Solving the two-level atom
Bloch equations in the weak excitation limit, and in steady state,
gives $\hat{\sigma}=i \sqrt{2\gamma_{\rm in}}\hat{a}_{\rm
in}/(\gamma+i\Delta)$, where $\Delta$ is the frequency detuning of
the probe from the excited state. The transmission of the
intensity of a probe field finally reads
\begin{eqnarray}\label{Eqabs}
T=|1-2\epsilon\mathcal{L}(\Delta)|^2,
\end{eqnarray}
where $\mathcal{L}(\Delta)=\gamma/(\gamma+i\Delta)$ for a two
level atom. This theory predicts full reflection of the probe
field for a weak resonant input field covering half of the full solid
angle. One important point here is the interference between the
incident beam and the radiated dipole field, which yields a
considerable decrease in the forward mode amplitude
\cite{Tey09,Zum08}. Using our numerical aperture NA=0.4 (i.e
$\epsilon=4\%$), we expect a probe beam extinction of $16\%$ using
Eq.~(\ref{Eqabs}). More refined models were proposed in
\cite{Venk04,Tey08,Zum08} using a decomposition into cylindrical
modes and including the dipole emission pattern. This set of modes
is adapted to the coupling of the input beam with a high numerical
aperture lens beyond the paraxial approximation. From the model of
\cite{Tey08}, we expect our extinction to be around $13\%$. We
note that for efficient information transfer between a single
photon and a single atom, i.e. efficient change of the atomic
states population by a single photon, the full dipole radiation
pattern and the reversed temporal mode of the atomic emission have
to be matched by the single photon input mode \cite{Son07}.

\begin{figure}[t]
\hspace*{-2mm}\centerline{\scalebox{0.27}{\includegraphics{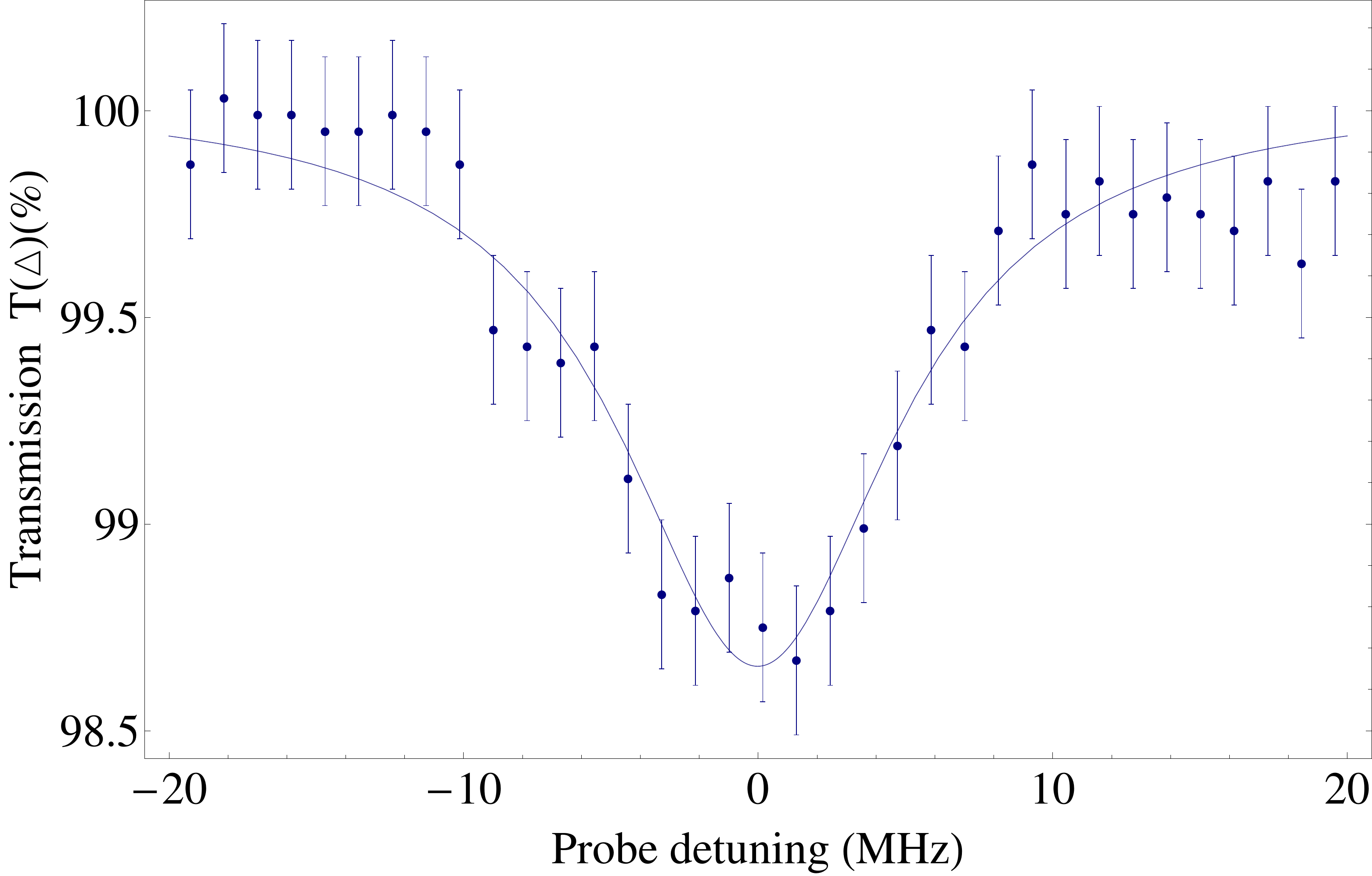}}}
\caption{Normalized power of a probe field transmitted through a
single trapped Barium ion as a function of probe beam detuning.
The transmission spectrum is fitted by a Lorentzian profile with a
width of 11 MHz. The peak probe beam extinction is 1.35\%.\\}
\label{absorption}
\end{figure}

Our experimental set-up and the level scheme of the atom are
depicted Fig.~\ref{setup}-(a),(c). We use a Barium ion as our
single-atom reflector. It is trapped and optically cooled in a
spherical Paul trap (with harmonic motion of about 1 MHz), using
two narrow-band and tunable red (649 nm) and green (493 nm) laser
fields driving the S$_{1/2}$ to P$_{1/2}$ and D$_{3/2}$ to
P$_{1/2}$ transitions, respectively. The ion is continuously
cooled on the S$_{1/2}$ to P$_{1/2}$ by the 493 nm laser that is
red detuned by 50 MHz with respect to resonance. The cooling beam
intensity is set far from saturation to minimize depopulation of
the $S_{1/2}$ state, yet allowing cooling to the Lamb-Dicke
regime. The saturation parameter obtained by fitting the four dark
resonances in the fluorescence spectrum is about $0.1$. Here, the
saturation parameter is defined for each transition as
$\Omega^2/(\gamma^2+\Delta^2)$, where $\Omega$, $\gamma$ and
$\Delta$ are the Rabi frequency, spontaneous decay rate and the
laser detuning of the particular transition, respectively. Atomic
population from the D$_{3/2}$ manifold is recycled by the red
{\it repumper no.\,1}, red detuned by 35 MHz and operated with a
saturation parameter of around $0.8$. For this configuration of
the laser intensities and frequencies simulations show, that in
steady state 70\,\% of the atomic population is in the
S$_{1/2}$(m=+1/2) Zeeman sublevel.

The weak probe beam is frequency-shifted with respect to the cooling beam by an
acousto-optic modulator and spatially filtered using a single-mode
fiber (SMF) to guarantee a Gaussian spatial profile. Its
polarization is adjusted by a quarter-wave plate
for efficient elastic scattering on the S$_{1/2}$(m=+1/2) to
P$_{1/2}$(m=-1/2) transition.

Another crucial part of the experiment is to overlap the incoming
probe beam with the dipole emission pattern. This mode-matching is
done using an expanding telescope and custom-designed objective
\cite{Linos} with a numerical aperture of 0.4. 1.5\,\% of the
transmitted probe together with a fraction of the ion's green
fluorescence is then collected by a microscope objective and
detected on photomultiplier PMT$_2$. The green fluorescence is
also detected in the backward direction by the PMT$_1$. A typical
fluorescence count rate measured on PMT$_1$ is 600 photons per
second, with the fluorescence from the probe beam contributing to
less than 100 photons per second. After considering detection
losses, these count rates are still by factor of more than ten
lower than the count rates typically observed for the S$_{1/2}$ to
P$_{1/2}$ transition saturation.

To allow for precise estimation of our extinction efficiency, we
modulate the repumper beam at 600\,Hz using a mechanical chopper (IM).
When the repumper beam is on, fast optical pumping to the S state
takes place, which allows both scattering of the probe and
cooling. When the repumper is off, however, the green cooling beam
depopulates the S state, so that the probe does not interact with
the ion. The probe signal intensity is then modulated at 600\,Hz
by the ion with a phase shifted by $\pi$ with respect to the
chopping signal. The signal from PMT$_2$ is subsequently
demodulated and low pass filtered. As a
first step, we find the local oscillator phase which yields the
maximum positive signal amplitude for the fluorescence
only, i.e. operating with large cooling field powers and with the
probe off. Next we turn down the cooling beam power to below
saturation in order to observe a negligible fluorescence signal
whilst still cooling the ion efficiently. With the probe field on,
reflection off the ion gives a negative signal (see
Fig.~\ref{setup}(b)), thus unambiguously discriminating the
fluorescence contribution from the probe extinction.

\begin{figure}[t]
\centerline{\scalebox{0.28}{\includegraphics{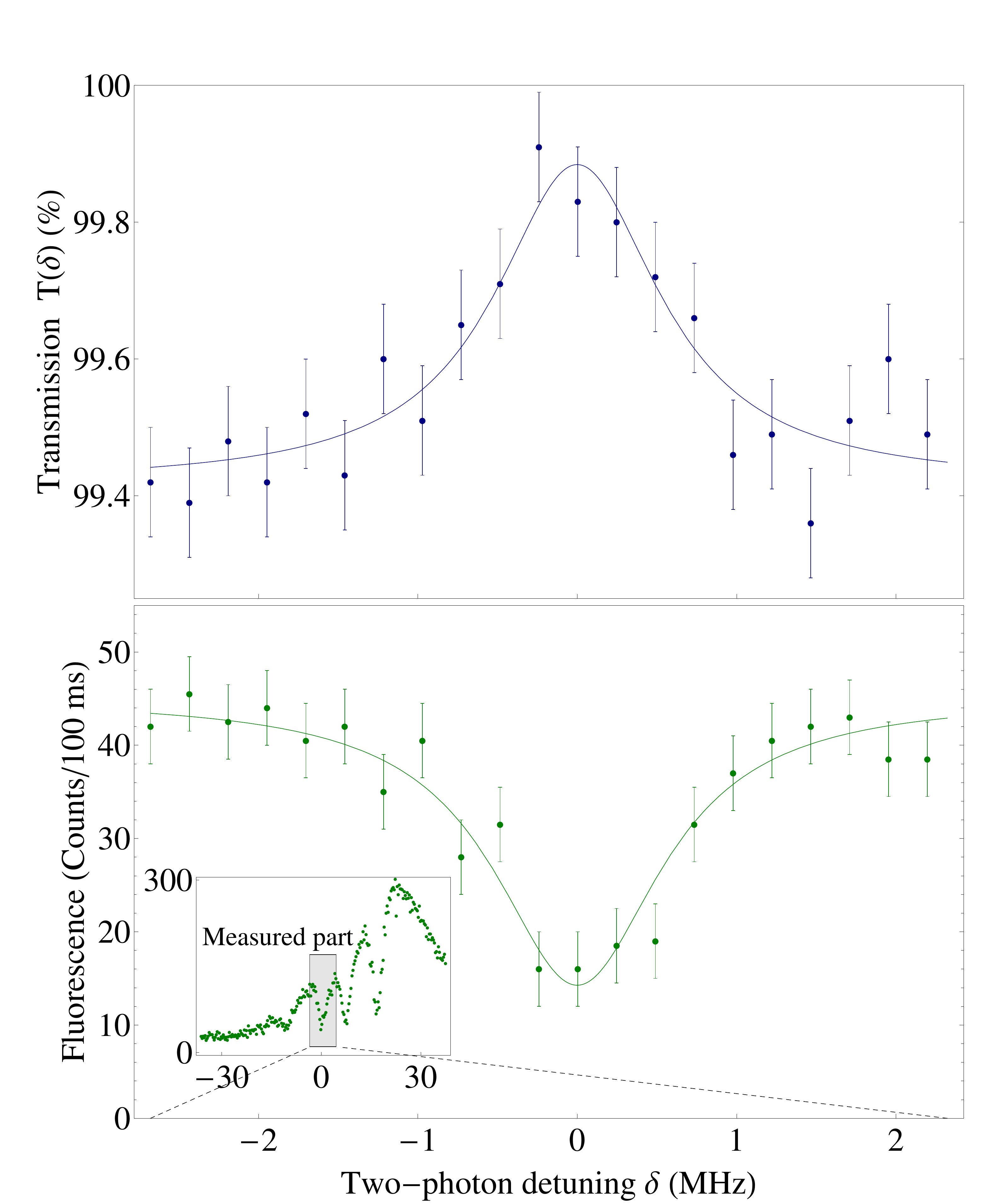}}\put(-248,285){a)}\put(-248,148){b)}}
\caption{Electromagnetically-Induced-Transparency (EIT) and
fluorescence spectra. a) Normalized transmitted power of a probe
field scattered by a single ion as a function of two-photon
detuning. The control beam frequency is tuned across a two-photon
resonance, which enhances the forward scattering via EIT. The
Lorentzian fit of the EIT transmission dip gives a linewidth of
1.2\,MHz and a suppression of the probe beam extinction by 75\,\%.
b) Dark resonance in the fluorescence spectrum simultaneously
measured in the backward direction. The inset shows a typical
fluorescence spectrum with all four dark resonances. The EIT
experiment was performed on the shaded part of the spectrum.}
\label{EIT}
\end{figure}

Fig.~\ref{absorption} shows a typical scan of the probe beam
extinction as a function of probe frequency. We observed a
Lorentzian dependence of the transmission profile with a width
of 11 MHz. A maximum of $1.35\,\%$ extinction was found on
resonance. The difference with the extinction predicted in
\cite{Tey09} can be partly explained by imperfect overlap of the
incoming probe polarization mode with the polarization of the
scattered light and by residual saturation of the S to P
transition by the cooling beam. Residual spherical
aberrations and atomic motion also reduce the spatial
matching of the probe with the dipole field.
Larger extinctions are likely
to be reached through better pumping preparation and by time
separation of the cooling and extinction measurement processes.

We now demonstrate electromagnetically induced transparency of the
ion on the probe using a dressing laser field on the red transition. Under
conditions of a weak probe and stronger control field, a narrow
transparency window 'opens' for the probe that would otherwise be
reflected in the absence of the control laser. This effect was
demonstrated by many groups using optically thick atomic ensembles
(see Ref. \cite{Bol91} for the first demonstration), but has not
been observed using single atoms in free space.

Under weak probe excitation, the probe transmission as a function
of the two-photon detuning $\delta=\Delta_g-\Delta_r$ can be found
by solving the Bloch equations \cite{Fle05} and using the above
input-output relations. Neglecting the angular dependence of the
extinction (due to polarization), we can replace the function
$\mathcal{L}$ by
\begin{eqnarray}\label{EITeq} \mathcal{L}_{\Lambda}(\delta)=\frac{\gamma(\gamma_0-i\delta)}{(\gamma_0-i\delta)(\gamma+i\Delta_g)+\Omega_r^2},
\end{eqnarray}
in Eq.~(\ref{Eqabs}), where $\Omega_r$ is the Rabi frequency of
the red laser field, $\gamma_0$ the ground state dephasing rate,
$\gamma$ the natural linewidth of the two transitions (assumed to
be the same for simplicity). An important condition for EIT to
take place is $\gamma\gamma_0 \ll \Omega_r^2$, i.e. the pumping
rate to the dark state must be much faster than any ground state
decoherence process. Independent frequency fluctuations of the two laser
fields, magnetic field fluctuations, and atomic motion induced Doppler
shifts, must be therefore reduced. When this is the case,
extinction of the resonant probe can be completely inhibited,
within a small range of control laser detuning
$\Omega_r^2/\gamma$, creating an EIT window. This is what we
observed in this experiment.

Here, we co-propagated the control and probe fields to eliminate
the effects of Doppler shifts due to the ion motion. We found that
the motion induced decoherence yields broadening of tens of kHz,
which reduced the EIT when the control and the probe were
orthogonal to each other. To optimize EIT conditions, we do
not use the cooling fields which would reduce the EIT process, so
the ion was now cooled by the probe itself. Consequently, a red
detuned and more intense probe was used, which gave extinction
efficiencies of about $0.6\%$. Due to the multi-level structure of
Barium, a single three level system can only be perfectly isolated
from the others through optical pre-pumping. Stark-shifts induced
by the other levels and double-$\Lambda$ type couplings here
contribute to a slight reduction of the EIT contrast.

Fig.~\ref{EIT}~a) shows the probe beam extinction strength as the
control field ({\it repumper no.\,2}) is scanned across the
two-photon resonance. We observe a large inhibition of the probe
beam extinction at zero two-photon detuning, with a peak value of
75\,\%. The measured linewidth of the EIT window is 1.2\,MHz, much
below the natural linewidth of the S to P transition.
Fig.~\ref{EIT}~b) shows the resulting scattered light intensity in
the backward direction, showing a corresponding decrease of the
fluorescence light around the two photon resonance, as expected
due to dark state pumping. Although the ion motion was too large
to yield significant extinction on the blue side, the other three
EIT profiles could also be observed. The contrast and width of
these two-photon resonances are mostly given by power broadening
and frequency fluctuations of our two lasers. The control and
probe laser linewidths are 80\,kHz and 20\,kHz respectively, which
allows a minimal EIT transmission linewidth of 82\,kHz\, to be
observed. Performing an efficient pre-pumping to the S state, and
switching the cooling fields off while the EIT is measured, would
allow us to reach ultra-narrow transmission profiles. This is of
significant interest as this means that large phase shifts are
imprinted on the probe field \cite{Bol91}, which might become
useful for precision spectroscopy and state detection with single
atoms.

In conclusion, we observed both the direct extinction of a weak
probe field and electromagnetically induced transparency from a
single Barium ion. The maximum observed extinction of the probe
beam intensity due to the scattering on the ion was 1.3\% and
inhibition of extinction due to EIT was almost 75\%. The amount of
extinction is limited mainly by the numerical aperture of the
employed lens, and residual saturation of the probed transition by
the cooling beams. Using a numerical aperture of say 0.7 and a
better pumping scenario, would already give extinction values of
up to 50\%, which are within experimental reach. Better EIT
contrasts and narrower features can be reached through efficient
preparation of an isolated $\Lambda$ scheme, with pre-pumping
steps.
Our results have a number of direct applications besides precision spectroscopy.
One can take
advantage of the sensitivity of EIT with regards to Doppler shifts
to read out the atomic motional state in a quantum non demolition
manner, thereby also opening the way towards quantum feedback
\cite{Rabl}. Furthermore, the presented results have direct
implications for long distance quantum information \cite{Cir97,DLCZ}
and quantum computation. Quantum memories, where quantum states
between atoms and light fields are reversibly exchanged, indeed
form an essential part of quantum repeater architectures
\cite{DLCZ,Fle05} and EIT is a prominent method to achieve such a
transfer \cite{Fle05}.

%\bibliography{EIT}

\end{document}